\documentclass[pra,twocolumn,floatfix,superscriptaddress,amsmath,citeautoscript,aps,longbibliography]{revtex4-2}
\usepackage{graphicx}  
\usepackage{dcolumn}   
\usepackage{bm}        
\usepackage{amssymb}   
\usepackage{amsmath}
\usepackage{color}
\usepackage{natbib}
\usepackage{hyperref}  
\hypersetup{hypertex=true,
            colorlinks=true,
            linkcolor=blue,
            anchorcolor=blue,
            citecolor=blue}

\newcommand{\TNA}{TmNi$_3$Al$_9$}

\begin{document}

\title{Interplay of itinerant electrons and Ising moments in a hybrid \\ honeycomb quantum magnet TmNi$_3$Al$_9$}

\date{\today}
\author{H. Ge}
\affiliation{Department of Physics, Southern University of Science and Technology, Shenzhen 518055, China}

\author{C.~J. Huang}
\affiliation{Department of Physics and HKU-UCAS Joint Institute
for Theoretical and Computational Physics at Hong Kong,
The University of Hong Kong, Hong Kong 999077, China}

\author{Q. Zhang}
\affiliation{Neutron Scattering Division, Oak Ridge National Laboratory, Oak Ridge, Tennessee 37831, USA}

\author{N. Zhao}
\affiliation{Department of Physics, Southern University of Science and Technology, Shenzhen 518055, China}

\author{J. Yang}
\affiliation{Department of Chemistry, Southern University of Science and Technology, Shenzhen 518055, China}

\author{L. Wang}
\affiliation{Department of Physics, Southern University of Science and Technology, Shenzhen 518055, China}
\affiliation{Shenzhen Institute for Quantum Science and Engineering,Shenzhen 518055, China}

\author{Y. Fu}
\affiliation{Department of Physics, Southern University of Science and Technology, Shenzhen 518055, China}

\author{L. Zhang}
\affiliation{Ganjiang Innovation Academy, Chinese Academy of Sciences, Ganzhou 341000, China}

\author{Z.~M. Song}
\affiliation{Department of Physics, Southern University of Science and Technology, Shenzhen 518055, China}

\author{T.~T. Li}
\affiliation{Department of Physics, Southern University of Science and Technology, Shenzhen 518055, China}

\author{F. Ding}
\affiliation{Department of Physics, Southern University of Science and Technology, Shenzhen 518055, China}

\author{J.~B. Xu}
\affiliation{Department of Physics, Southern University of Science and Technology, Shenzhen 518055, China}

\author{Y.~F. Zhang}
\affiliation{Department of Physics, Southern University of Science and Technology, Shenzhen 518055, China}

\author{X. Tong}
\affiliation{Institute of High Energy Physics, Chinese Academy of Sciences, Beijing 100049, China}
\affiliation{Spallation Neutron Source Science Center, Dongguan 523803, China}

\author{S.~M. Wang}
\affiliation{Department of Physics, Southern University of Science and Technology, Shenzhen 518055, China}

\author{J.~W. Mei}
\affiliation{Department of Physics, Southern University of Science and Technology, Shenzhen
  518055, China}
\affiliation{Shenzhen Institute for Quantum Science and Engineering,Shenzhen
  518055, China}
\affiliation{Shenzhen Key Laboratory of Advanced Quantum Functional Materials
   and Devices, Southern University of Science and Technology, Shenzhen 518055, China}

\author{A. Podlesnyak}
\affiliation{Neutron Scattering Division, Oak Ridge National Laboratory, Oak Ridge, Tennessee 37831, USA}

\author{L.~S. Wu}
\email{wuls@sustech.edu.cn}
\affiliation{Department of Physics, Southern University of Science and Technology, Shenzhen 518055, China}
\affiliation{Shenzhen Key Laboratory of Advanced Quantum Functional Materials and Devices,
Southern University of Science and Technology, Shenzhen 518055, China}

\author{Gang Chen}
\email{gangchen@hku.hk}
\affiliation{Department of Physics and HKU-UCAS Joint Institute for Theoretical and Computational Physics at Hong Kong,
The University of Hong Kong, Hong Kong 999077, China}

\author{J.~M. Sheng}
\email{shengjm@sustech.edu.cn}
\affiliation{Academy for Advanced Interdisciplinary Studies, Southern University of Science and Technology,
Shenzhen 518055, China}
\affiliation{Department of Physics, Southern University of Science and Technology,
Shenzhen 518055, China}

\date{\today}

\begin{abstract}
The interplay between itinerant electrons and local magnetic moments in quantum materials
brings about rich and fascinating phenomena and stimulates various developments in the theoretical framework.
In this work, thermodynamic, electric transport, and neutron diffraction measurements were performed on a newly synthesized honeycomb lattice magnet TmNi$_3$Al$_9$.
Based on the experimental data, a magnetic field temperature phase diagram was constructed, exhibiting three essentially different magnetic regions.
Below ${T_{\rm N}=2.97 \pm 0.02}\ \rm K$ Tm$^{3+}$ moments order antiferromagnetically in zero field.
We found that the Tm$^{3+}$ ions form a pseudo-doublet ground state with the Ising-like moments lying normal to the two-dimensional honeycomb layers.
Application of a magnetic field along the easy axis gradually suppresses the antiferromagnetic order in favor of an induced ferromagnetic state above the critical field ${B_c=0.92 \pm 0.05}\ \rm T$.
In the vicinity of $B_c$, a strong enhancement of the quantum spin fluctuations was observed.
The quantum Ising nature of the local moments and the coupling to itinerant electrons are discussed.
\end{abstract}

\maketitle

\section{INTRODUCTION}

Itinerant frustration occurs often in hybrid quantum materials with both itinerant electrons and local moments.
The frustration can come from the exchange interaction between local moments, or it can be generated from the Ruderman-Kittel-Kasuya-Yosida (RKKY) interaction via itinerant electrons~\cite{RKKY1,RKKY2}.
Frustrated interactions of local moments are among the most well-studied and attractive models in solid-state physics.
The frustration refers to the presence of localized magnetic moments interacting through competing exchange interactions, which results in a large degeneracy of the system ground state.
Frustration often leads to the formation of spin liquids ~\cite{LBalents2010,LSavary2017,ZhouYi2017}, in which spins fluctuate strongly down to a temperature of absolute zero.

When the quantum phases and the magnetic orders of the local moments feedback to the itinerant electrons, more interesting physics can emerge~\cite{ItinerantFrus}.
The magnetic order can reconstruct the band structures of itinerant electrons and generate nontrivial band-structure topology such as the well-known magnetic Weyl semimetal~\cite{MagWeylSemi1,MagWeylSemi2,MagWeylSemi3}.
The magnetic fluctuations can influence the physical properties of itinerant electrons and often convert
them into non-Fermi-liquids with distinct transport behaviors~\cite{TransBehaviors1,TransBehaviors2,TransBehaviors3,TransBehaviors4}.
The nature of the magnetic transition is further modified by gapless itinerant fermions.
Moreover, the manipulation of one degree of freedom can influence the physical properties of the other degree of freedom.
Thus, the interplay between itinerant electrons and local moments in hybrid quantum materials could generate rather rich phenomena for both fundamental research and application purposes~\cite{Application}.

Here, we employed magnetization, specific heat, electrical transport, and neutron diffraction measurements to study the physical properties of the two-dimensional (2D) honeycomb magnet \TNA\ in detail, and we constructed a magnetic field temperature phase diagram. \TNA\ has both local rare-earth moments and conduction electrons, which makes this material a model system in which to examine the effect of the coupling of multiple degrees of freedom on transport and magnetic properties.
We found that the magnetic properties of \TNA\ are highly sensitive to the magnetic field.
By applying the field along the easy $c$~axis, the magnetic order is gradually suppressed to zero temperature at the critical field ${B_c=0.92 \pm 0.05}\ \rm T$.
Strong quantum critical fluctuations were observed in the vicinity of the quantum critical field.
This is further reflected in the anomalous scaling of the electric resistivity in transport measurements.

\section{EXPERIMENTAL DETAILS}

High-quality~\TNA\ single crystals are grown using the metallic flux method~\cite{TY2011}.
The crystal structure of~\TNA\ is characterized using a Bruker D8 Quest diffractometer with Mo-K$_{\alpha}$ radiation (${\lambda=0.71073~\rm\AA}$). The commercial Bruker APEX2 software suite was used for data integration and reduction. To identify the magnetic structure, neutron
powder diffraction measurements with center wavelengths 1.5 ${\rm\AA}$ were performed with the high-resolution time-of-flight powder diffractometer POWGEN, at the Spallation Neutron Source, Oak Ridge National Laboratory~\cite{Huq2011}. A powder sample of~\TNA, prepared by grinding about $2~g$ of single crystals, was loaded in vanadium can for the measurements. An orange cryostat
was used to cover the temperature region from 1.7 K to 300 K. The magnetic structure refinement was carried out
using the software package FullProf~\cite{Rod1993}.
Specific heat and electrical resistivity measurements were performed on the Physical Property Measurement System. Magnetization measurements were performed using two magnetometers for different temperature ranges.
For $T=1.8 - 300$~K, a commercial Quantum Design Magnetic Property Measurement System was used, while for $T=0.4 - 1.8$~K, a Hall sensor magnetometer integrated in the $^{3}$He insert was used~\cite{QDMagnetometry,ACandini2006,ACavallini2004}.

\section{Results and Analysis}

\subsection{Crystal Structure and Crystal Electric Field}

\begin{figure}[tbh]
 \includegraphics[width=0.75\linewidth]{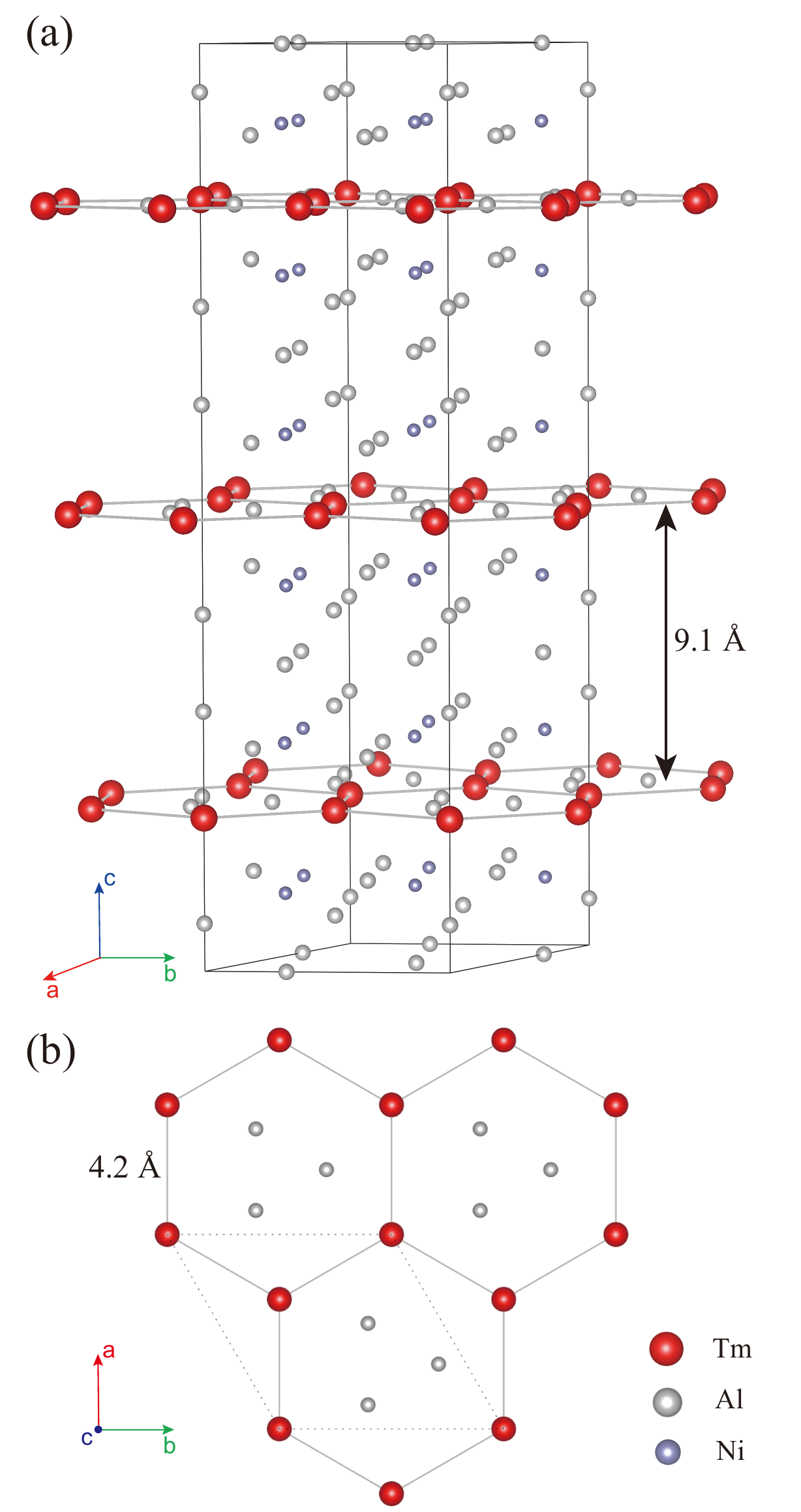}
    \caption{(a) The sketch of the crystal structure of TmNi$_3$Al$_9$. (b) Schematic view of a single 2D honeycomb layer in $ab$ plane.}
\label{fig:structure}
\end{figure}

As a first step, we performed a characterization of our single crystals using single crystal x-ray measurements.
The room temperature lattice parameters, $a=b=7.2576\ {\rm\AA},c=27.2983\ {\rm\AA}$ and $\alpha=\beta=90^{\rm\circ},\gamma=120^{\rm\circ}$, are consistent with previous report~\cite{TY2011}.
In this ErNi$_3$Al$_9$-type structure (the trigonal $R32$ space group), the honeycomb lattice of Tm$^{3+}$ ions within crystallographic $ab$ planes is well separated along the $c$ axis.
The inter-layer spacing is about 9.1~\AA, in comparison with the value of 4.2~\AA\ observed for the in~plane nearest-neighbor (NN) distance [see Fig.~\ref{fig:structure}(a,b)].

Due to the large spin-orbit coupling of 4$f$ electrons, the crystal electric field (CEF) is important for understanding the magnetism of rare-earth ions.
The 3-fold point symmetry of the Tm site indicates that the Tm$^{3+}$ magnetic moments are constrained either in the $ab$ plane (perpendicular to the high symmetric direction), or along the $c$ axis (parallel to the high symmetric direction).
Further calculations of the ground state wave functions and the CEF configuration were performed using the software package McPhase~\cite{MCPHASE2004, PointCharge1964}.
In this local environment, the 13-fold degenerate ${J = 6}$ (${L = 5, S = 1}$) multiplet (${2J + 1 = 13}$) of $\rm Tm^{3+}$ is split into several doublet and singlet states.
Since the Kramers theorem does not apply here, there is no guarantee for a degenerate doublet ground state.
However, by choosing the high symmetry axis $c$ as the local $z$ direction, our point charge model calculations indicate a pseudo doublet state with the wave functions mostly contributed from ${|\pm6\rangle}$.
This is similar to the case of the triangular lattice antiferromagnet TmMgGaO$_4$~\cite{TMGO1,TMGO2}.
Thus, we expect magnetic moments with the local easy direction along the $c$ axis, which are Ising-like with the biggest contribution from the wave function ${|\pm6\rangle}$.

\subsection{Magnetization and Single Ion Anisotropy}

\begin{figure}[tbh]
\includegraphics[width=0.8\linewidth]{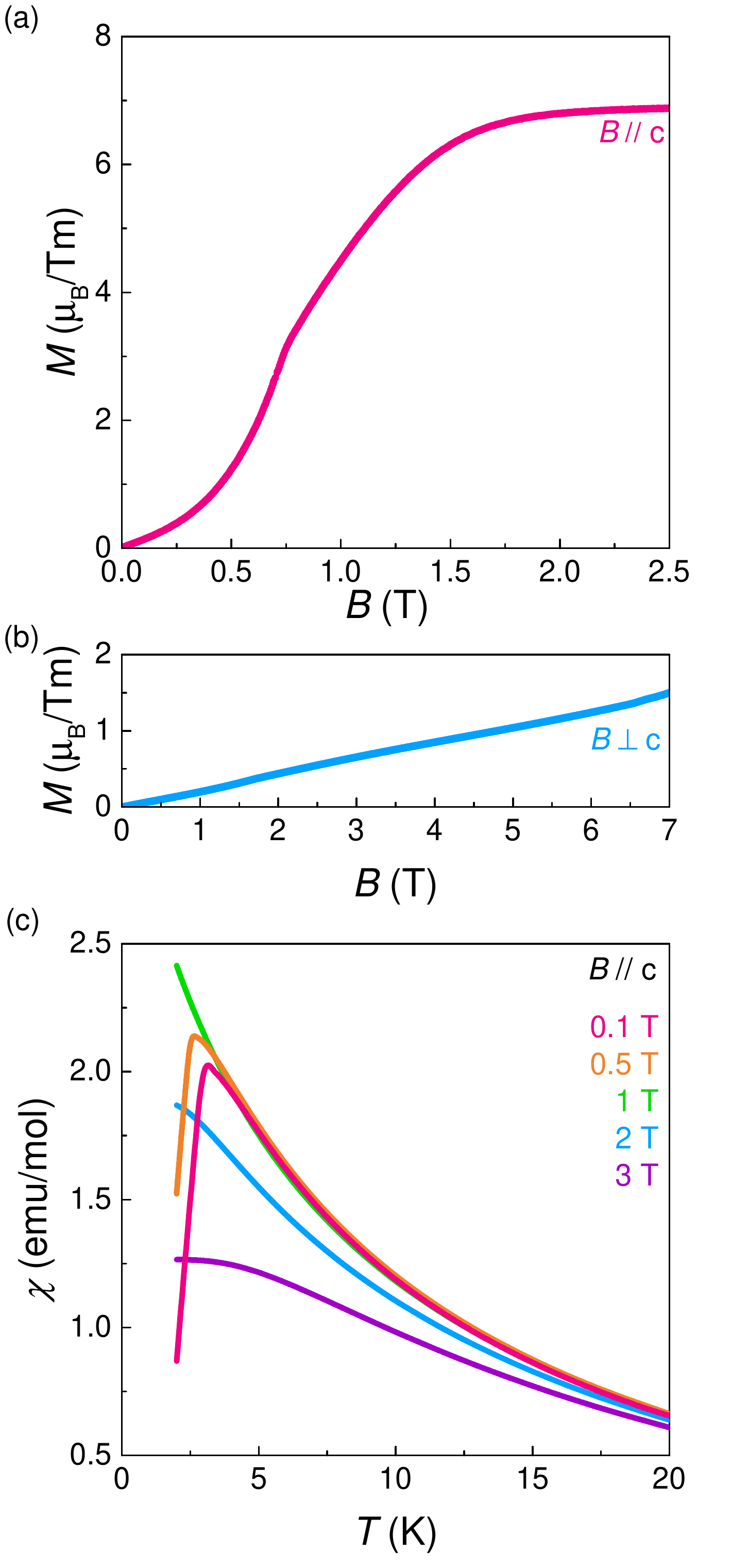}
\caption{Field-dependent magnetization $M$ of TmNi$_3$Al$_9$, measured at $T = 1.8$ K, with the field (a) $B\parallel c$ and (b) $B\perp c$. (c) Temperature-dependent DC magnetic susceptibility $\chi\rm$ at different magnetic fields for $B\parallel c$.}
\label{fig:magnetization1}
\end{figure}

The field-dependent magnetization of TmNi$_3$Al$_9$ measured at ${1.8}$ K with the magnetic field applied in the $ab$ plane and along the $c$ axis, are presented in Fig.~\ref{fig:magnetization1}(a,b).
The obtained results show significant anisotropic behavior.
For $B\parallel c$, the Tm$^{\rm 3+}$ moments are polarized above 2 T, with the saturation moment ${M_{\rm s}\simeq6.97~\mu_{\rm B}}$. This is very close to the expected value for pure ${|\pm6\rangle}$ ground states, where ${M_{\rm s} = g_{\rm J}J\mu_{\rm B}
=  7/6\times 6~\mu_{\rm B} = 7~\mu_{\rm B}}$.
The in-plane $B \bot c$ magnetization is rather different.
Linear field dependence is observed up to the highest measured field, $B=7$~T.
The value of the $B \bot c$  magnetization at 7 T is about one order of magnitude smaller than the saturation moment with the field along the $c$ axis.
This is consistent with the CEF calculations, and it supports our suggestion that the magnetic Tm$^{\rm 3+}$ moments are Ising-like, with an easy axis lying along the $c$~axis.

\begin{figure}[t]
 \includegraphics[width=0.9\linewidth]{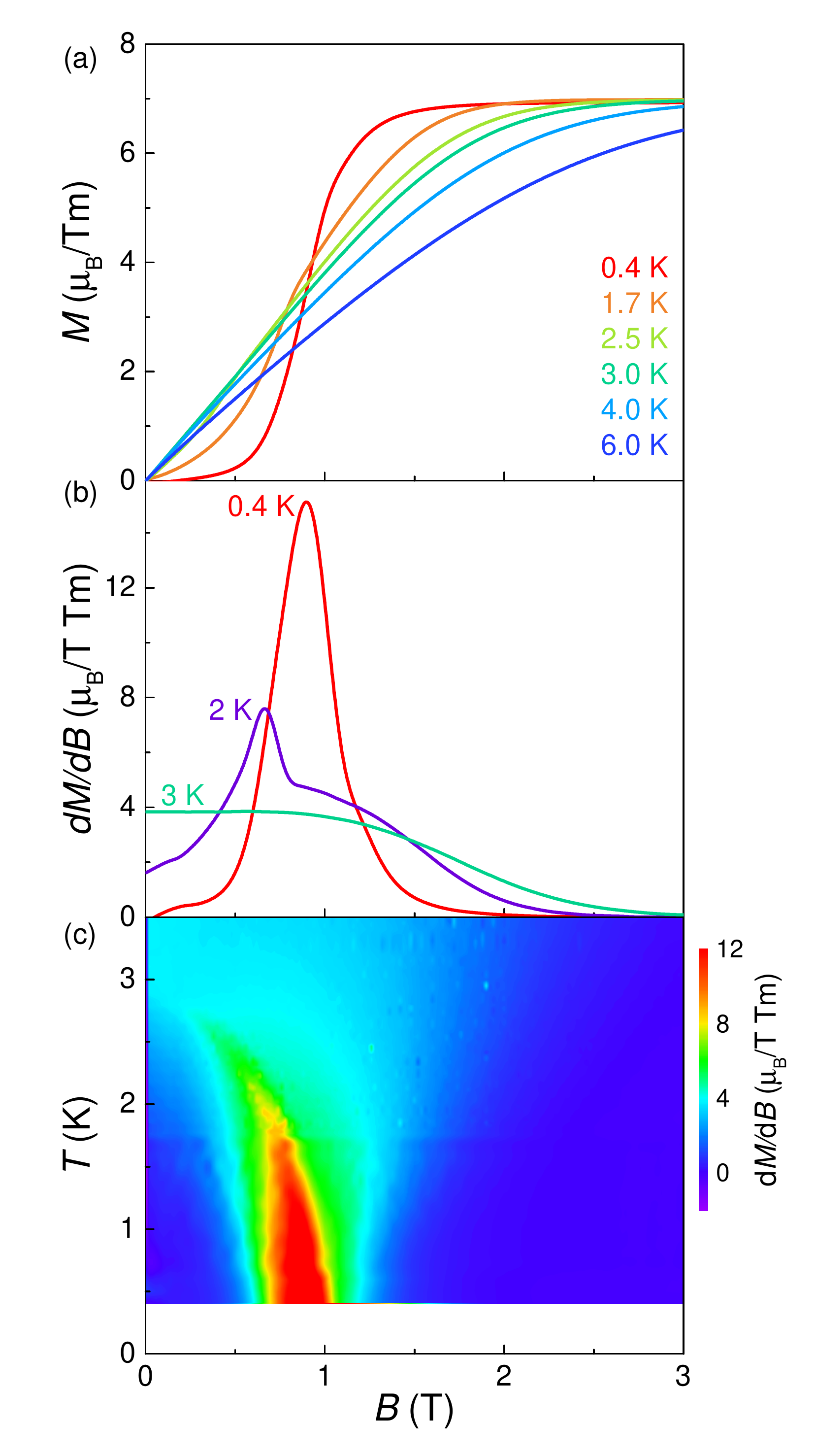}
    \caption{(a) Temperature dependence of the magnetization, measured at different temperatures for $B\parallel c$. (b) Magnetic field dependence of the deferential susceptibility $dM/dB$ at three selected temperatures, 3 K, 2 K, and 0.4 K. (c) The contour plot of the deferential susceptibility $dM/dB$.}
\label{fig:magnetization2}
\end{figure}

The temperature evolution of the DC magnetic susceptibility $\chi(T)$ with the magnetic field applied along the $c$ axis is shown in Fig.~\ref{fig:magnetization1}(c). Cusp-like anomalies are observed, indicating the establishment of a long-range antiferromagnetic (AFM) state. In a small field of 0.1 T, the magnetic transition is found at ${T_{\rm N}\simeq3}$ K. This transition is gradually suppressed upon increasing the field up to 1 T. Further magnetic characterization for temperatures lower than 1 K is performed on a home-built magnetometer. Shown in Fig.~\ref{fig:magnetization2}(a) is the field dependence of the magnetization, measured at different temperatures from 0.4 K to 6 K. The corresponding derivative susceptibility $dM/dB$ is presented in Fig.~\ref{fig:magnetization2}(b,c).
Below $T_{\rm N}$, peak-like anomalies in $dM/dB$ are observed, indicating the phase transition from the AFM order to the high field polarized state.

The N\'{e}el temperature is continuously suppressed to zero at a magnetic field-induced quantum critical point (QCP), at a critical field ${B_c=0.92 \pm 0.05}\ \rm T$.
We also noticed that in the intermediate temperature range ($1$-$2$ K) a broad shoulder-like anomaly is observed following the peak in $dM/dB$.
This shoulder-like feature then gradually merges with the peak at lower fields as further decreasing temperatures and finally evolves into a broad peak at a base temperature ${\sim0.4}$ K, indicating that strong spin fluctuations persist near the quantum critical field.

\subsection{Neutron Diffraction and Magnetic Structure}

\begin{figure}[tbh]
 \includegraphics[width=0.95\linewidth]{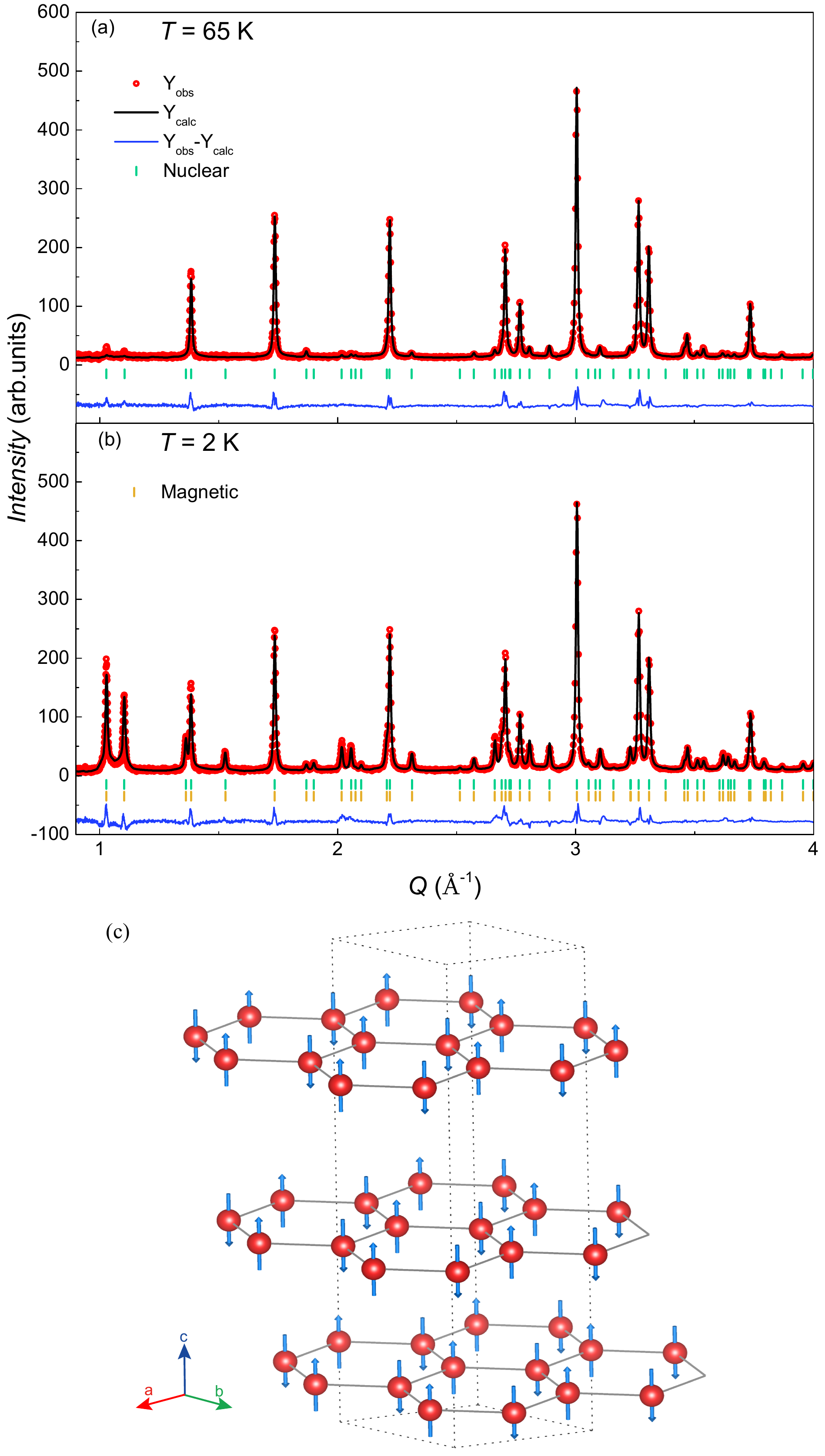}
    \caption{The results of Rietveld refinement of the neutron powder diffraction patterns of TmNi$_3$Al$_9$ at (a) 65 K and (b) 2 K.
    The red points represent the experimental data and black line is the model profile. The difference curve is shown at the bottom.
    (c) Schematic visualization of the magnetic structure for the magnetic phase of TmNi$_3$Al$_9$.}
\label{fig:magneticstructure}
\end{figure}

To determine the spin configurations in the AFM ordered state, neutron diffraction experiments are performed on powdered~\TNA\ samples.
The neutron diffraction patterns measured at 65~K (above $T_{\rm N}$) and 2~K (below $T_{\rm N}$) are shown in Fig.~\ref{fig:magneticstructure}(a) and (b), respectively.
Rietveld analysis confirms the trigonal structure with the space group $R32$, consistent with previous XRD results. 
For $T=2$ K, extra reflections emerge, especially around the low $Q$ region, suggesting the magnetic origin of these peaks.

\begin{figure}[tbh]
 \includegraphics[width=0.8\linewidth]{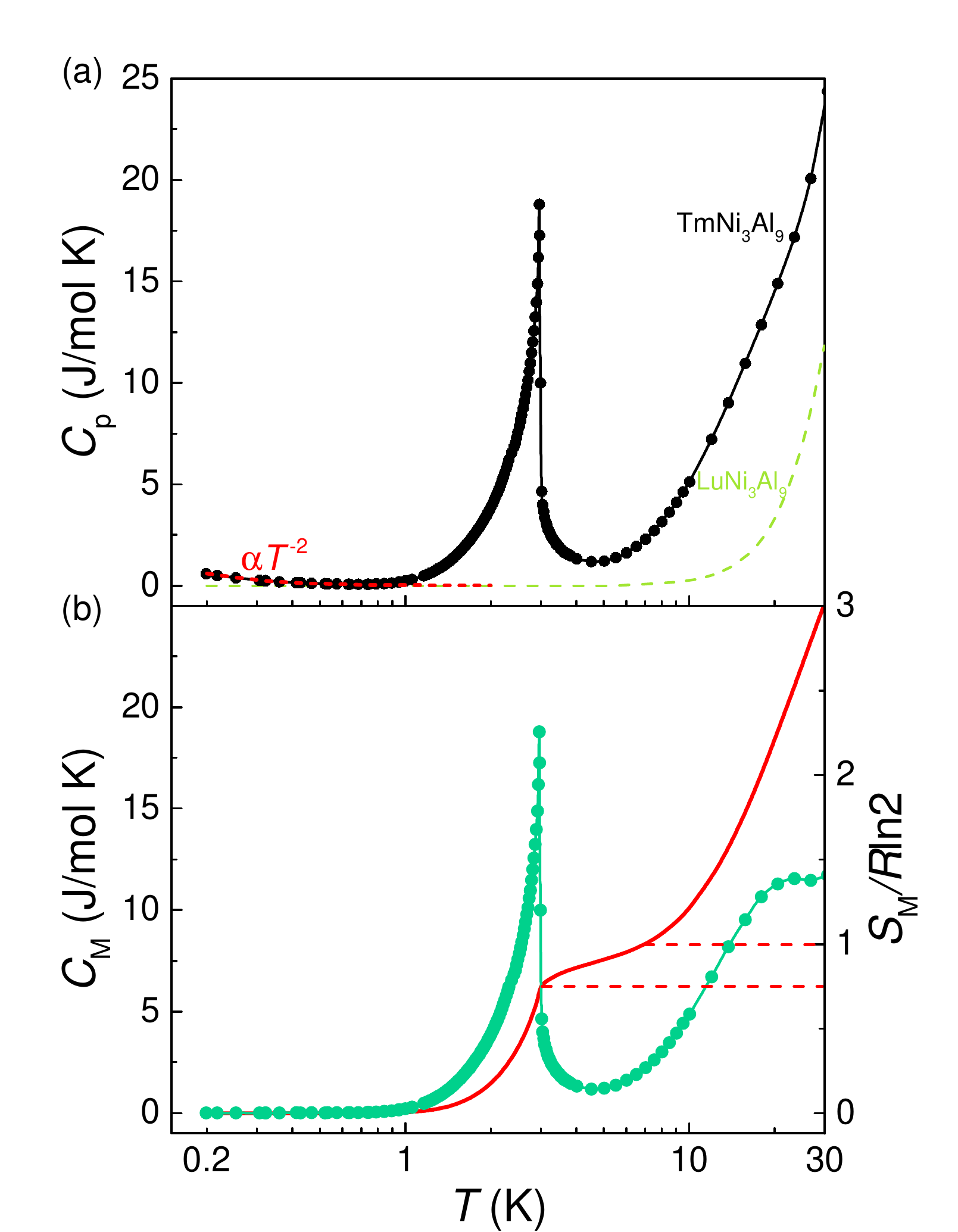}
    \caption{(a) Temperature dependence of the zero-field specific heat $C_p$ for TmNi$_3$Al$_9$ (solid black circles) and LuNi$_3$Al$_9$ (dashed green line). The red dashed line is the estimated contribution from the nuclear Schottky effect. (b) The magnetic specific heat $C_M$ and temperature-dependent entropy (red line), normalized to $R\ln$2, assuming a doublet ground state.}
\label{fig:heatcapacity}
\end{figure}

No additional (structural and/or magnetic) reflections appear below the magnetic transition.
The increase in the intensity of nuclear Bragg peaks due to magnetism is noticeable, especially in the low $Q$ region.
This implies that the magnetic propagation vector is ${\textbf{K}=\textbf{0}}$.
Symmetry-allowed magnetic space groups are analyzed using the Bilbao crystallographic server~\cite{BilbaoCrystal1,BilbaoCrystal2,BilbaoCrystal3,BilbaoCrystal4,BilbaoCrystal5,BilbaoCrystal5}. Two maximal magnetic space groups with nonzero magnetic moments, $R32' (\#155.47)$  and  $R32 (\#155.45)$ are possible. The magnetic space group $R32' (\#155.47)$ corresponds to a ferromagnetic spin configuration, which contradicts the magnetization measurements. In contrast, the magnetic space group $R32 (\#155.45)$ corresponds to the AFM structure with spins constrained along the $c$~axis. By refining both nuclear and magnetic reflections at 2 K, the lattice and magnetic structures are obtained simultaneously. The refinement confirms a $G$-type AFM configuration, where the spins are aligned antiferromagnetically in all $a$, $b$ and $c$ directions, as shown in Fig.~\ref{fig:magneticstructure}(c). The ordered $\rm Tm^{3+}$ moments are along the $c$ axis, consistent with the CEF calculation and the magnetic susceptibility measurement. The ordered moment of Tm$^{3+}$ ion is about $5.53(3)~\mu_{\rm B}$, smaller than the saturated moment $7~\mu_{\rm B}$. This is attributed to the quantum fluctuation generated by the intrinsic transverse field due to the weak splitting of the doublets, which is analogous to TmMgGaO$_4$~\cite{TMGO1,TMGO2}.

\subsection{Specific Heat and Magnetic Entropy}

\begin{figure}[tbh]
\includegraphics[width=0.8\linewidth]{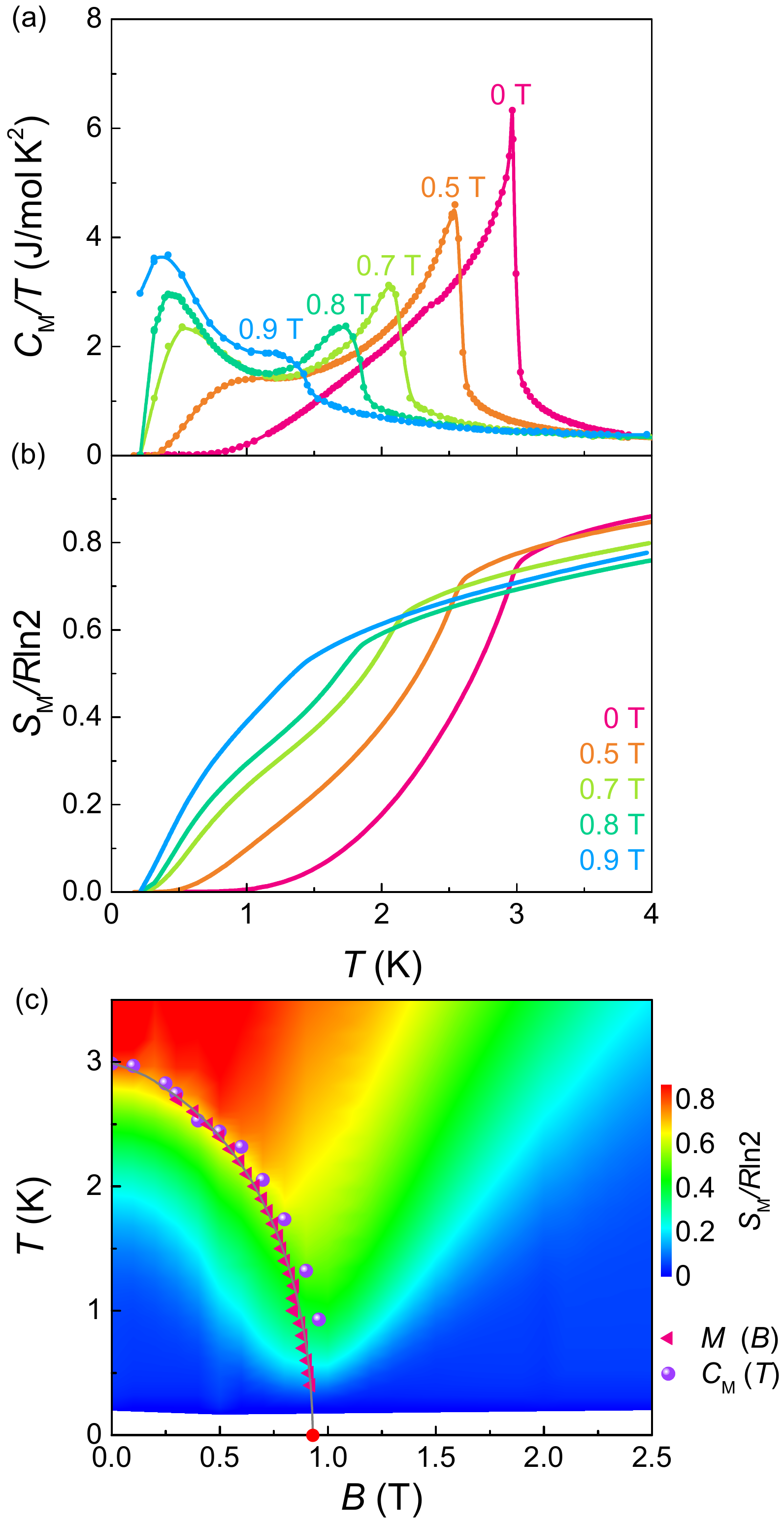}
    \caption{ The magnetic specific heat $C_{M}/T$ (a) and the integrated magnetic entropy $S_M$ (b) as a function of temperature at different magnetic fields. (c) The contour plot of the integrated magnetic entropy $S_M$. The triangles (circles) are data points extracted from peak positions of the magnetization (magnetic specific heat).}
\label{fig:Sm}
\end{figure}

To further investigate the spin fluctuations in~\TNA\ we performed specific heat measurements in different magnetic fields.
The zero-field specific heat is shown in Fig.~\ref{fig:heatcapacity}(a). The specific heat of an isostructural nonmagnetic compound LuNi$_3$Al$_9$ (Ref.~\cite{TY2011}) is used as the estimation of the lattice contribution (green dashed line in Fig.~\ref{fig:heatcapacity}(a)). A weak upturn of ${C_{\rm n}\sim \alpha T^{-2}}$ is observed below 0.5 K, due to the nuclear Schottky effect.
With the lattice and nuclear contribution removed, the magnetic specific heat exhibits a sharp peak at ${T_{\rm N}=2.97 \pm 0.02}\ \rm K$ that implies long-range AFM order (Fig.~\ref{fig:heatcapacity}(b)).
The integrated magnetic entropy reaches a value of $R\ln2$ at $\sim$7~K, consistent with the expectation for the pseudo doublet state of Tm$^{3+}$.
The further increase in temperature causes an additional magnetic contribution due to the thermal population of the excited CEF levels.

\begin{figure}[tbh]
 \includegraphics[width=0.8\linewidth]{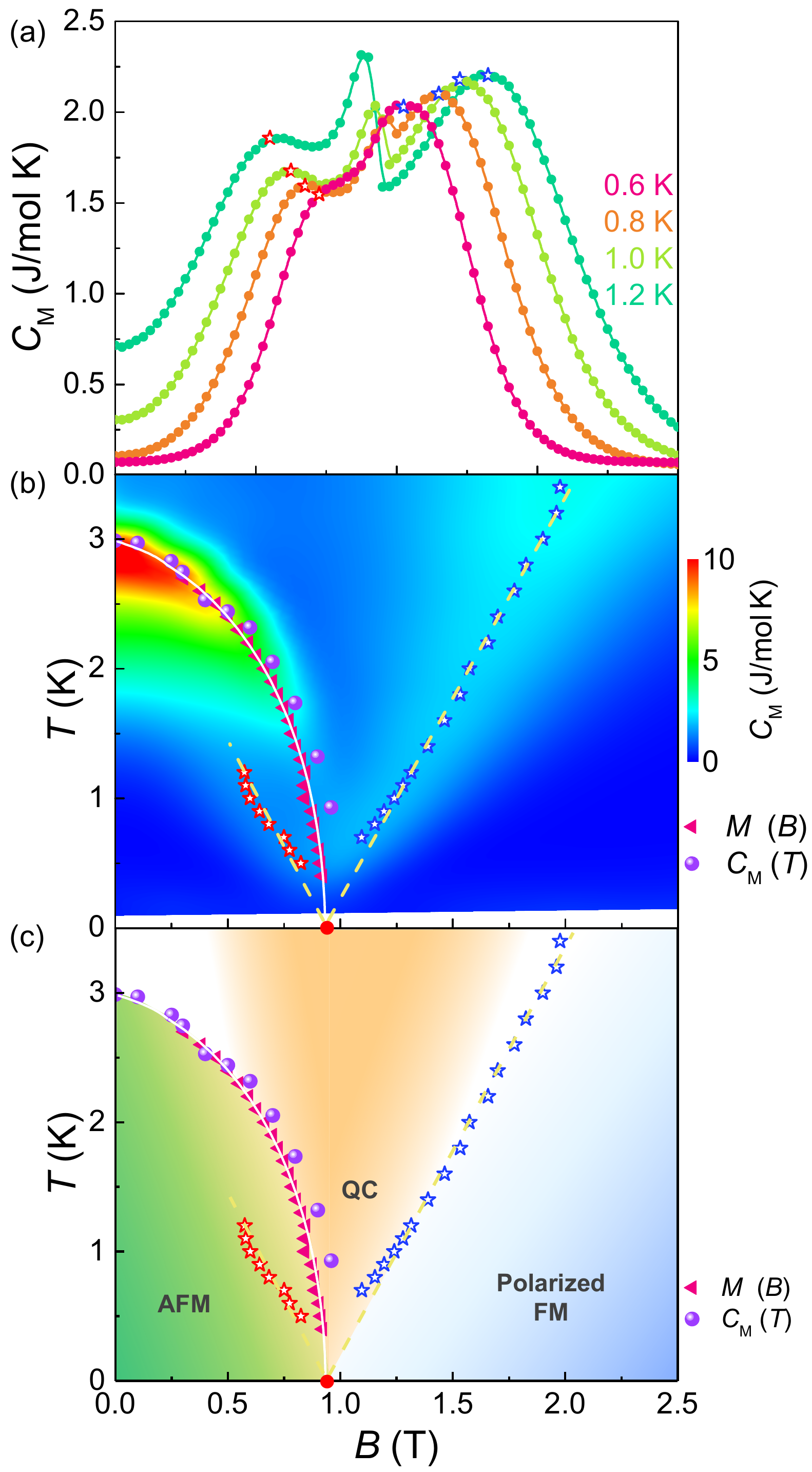}
    \caption{(a) Field dependence of the specific heat at different temperatures. (b) The contour plot of the magnetic specific heat of \TNA. The solid circles are the phase boundaries determined through the sharp peak positions of the specific heat, and the open stars are the crossover extracted from the broad peak positions as indicated in (a).
    (c) Phase diagram of \TNA\ reconstructed from the specific heat and magnetization measurements for $B\parallel c$. Colored areas, obtained from the integrated magnetic entropy, indicate long-range AFM, quantum critical, and field-polarized regimes. The red circle denotes the QCP at the critical field $B_c=0.9$~T.}
\label{fig:HC_phasediagram}
\end{figure}

The temperature dependence of the heat capacity at various fields is presented in Fig.~\ref{fig:Sm}(a).
With increasing field, the magnetic transition is gradually suppressed.
It is worth noting that the sharp peak observed in zero field, gets lower and broader with increasing magnetic field.
In the magnetic field above 0.7~T, an additional broad shoulder-like anomaly develops at lower temperatures.
The integrated magnetic entropy at different fields is shown in Fig.~\ref{fig:Sm}(b,c).
In zero field, about $80\%$ of $S_M / R\ln2$ is released at the transition temperature.
However, this ratio decreases at the elevated magnetic field.
Only about $50\%R\ln2$ is reached at the transition temperature in a field 0.8~T.
These phenomena suggest that the spin fluctuations are enhanced when the system is tuned to the critical region.

We summarize the $B-T$ phase diagram of \TNA\ in Fig.~\ref{fig:HC_phasediagram}(c), using the result of specific heat and magnetization measurements.
Field dependence of the magnetic specific heat reveals a step-like anomaly at high temperatures across the phase boundary as illustrated in Fig.~\ref{fig:HC_phasediagram}(a).
The anomaly gradually evolves into a weak peak at the critical field $B_c=0.9$~T.
This is due to the constraint of the third law of thermodynamics, where the entropy change ${\Delta S\rightarrow0}$ as ${T\rightarrow0}$ [Fig.~\ref{fig:HC_phasediagram}(b)].
In addition, crossover-like anomalies appear symmetrically around the critical field, as indicated by the open stars in Fig.~\ref{fig:HC_phasediagram}.
These crossover-like peaks are most distinguishable in the temperature range around 0.8 - 1.2~K, and then they gradually merge into one single peak at the critical field at temperatures below $\sim$0.6 K.
This behavior is rarely observed in traditional Ising magnets, where first-order classical spin-flip transition usually occurs.
Instead, the linear crossovers are typically expected in low-dimensional quantum magnets, where a number of spin fluctuations are present in the vicinity of QCP \cite{OBreunigSA2017,BYangPRL2017,ZWangPRL2018,LSWuNC2019, SFriedemannNP2009,PMerchantNP2014,SGJungNC2018,RHentrichPRL2018,BShenNature2020}.

\subsection{Electrical Transport and Non-Fermi Liquid Like Behaviors}

\begin{figure}[tbh]
 \includegraphics[width=0.8\linewidth]{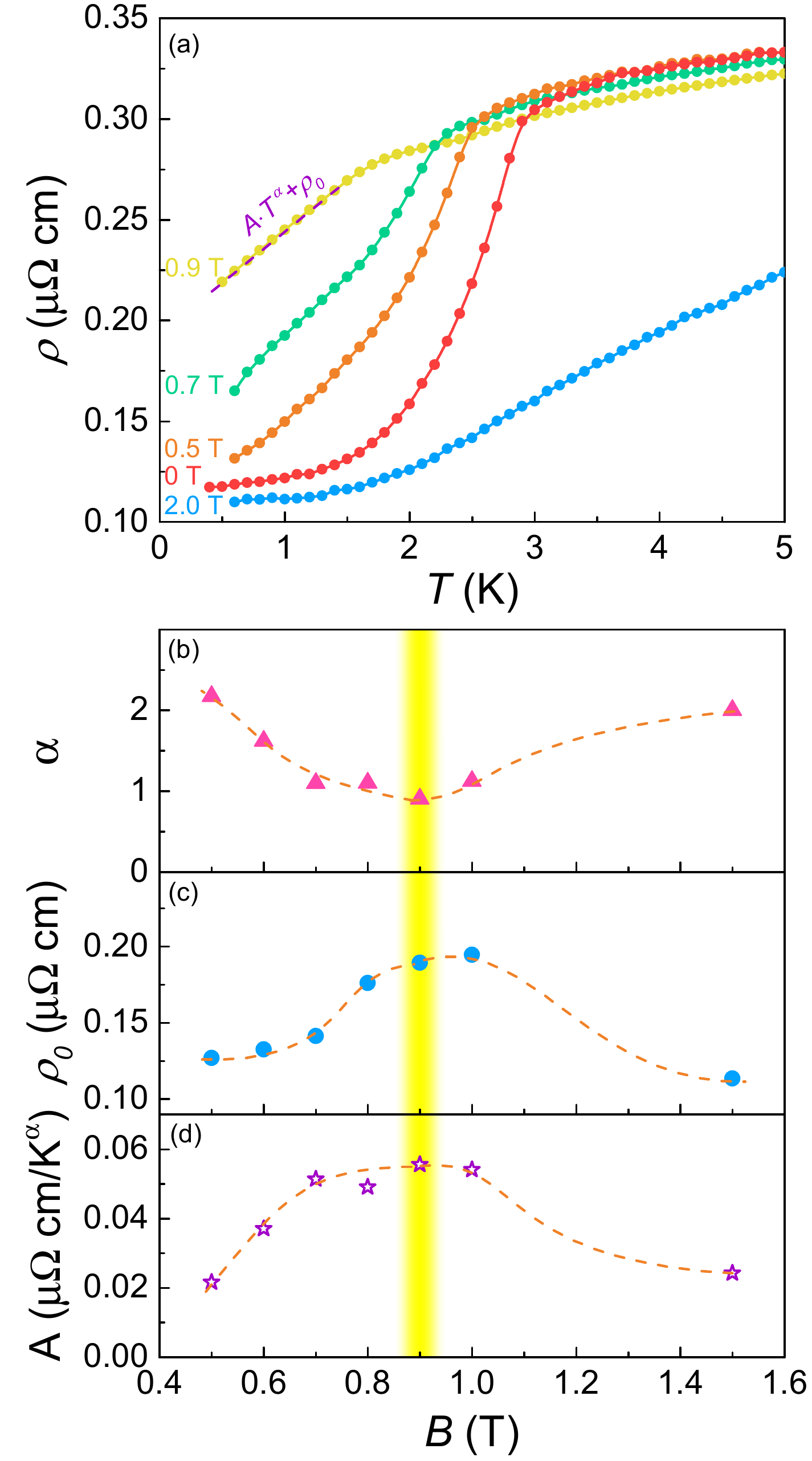}
    \caption{(a) Temperature dependence of the electrical resistivity of TmNi$_3$Al$_9$ at the various magnetic fields. (b)-(d) Magnetic field dependencies of the coefficients $\alpha$, $\rho_0$, and $A$, extracted from the fitting of $\rho=A\cdot{T^{\alpha}}+{\rho_{0}}$. Orange dashed lines are a guide to the eyes, and the yellow area indicates the region near the quantum critical point.}
\label{fig:RT}
\end{figure}

\begin{figure}[tbh]
 \includegraphics[width=0.95\linewidth]{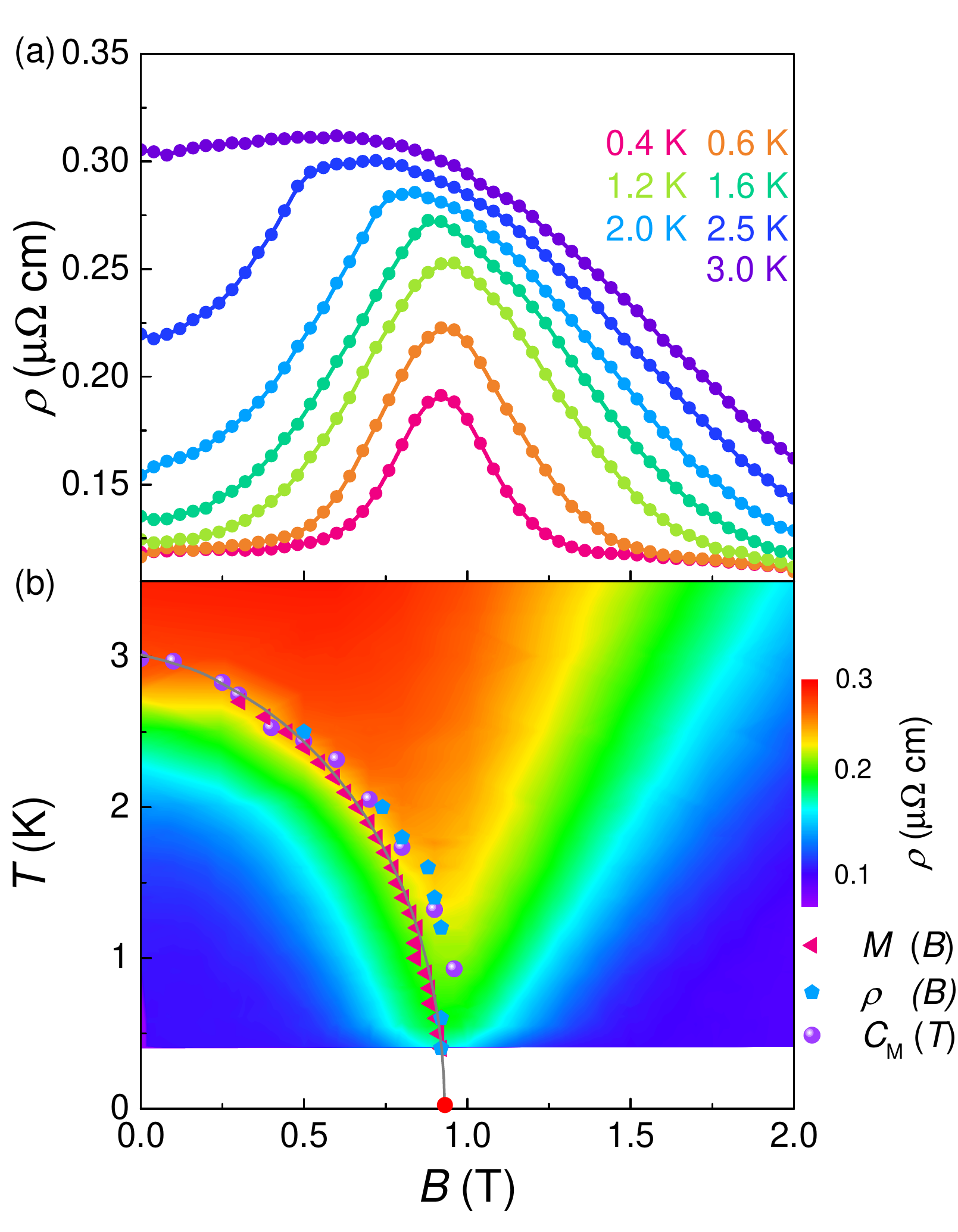}
    \caption{(a) Field dependence of the electrical resistivity with field applied along the $c$ axis. (b) The  contour plot of the resistivity at different temperatures and magnetic fields.}
\label{fig:RB}
\end{figure}

To reveal the impact of the local moment on itinerant electrons, the electrical resistivity of~\TNA\ with field along the crystal $c$ axis is performed as well. The temperature-dependent resistivity $\rho (T)$ is shown in Fig.~\ref{fig:RT}(a) at various fields from 0 T to 2 T. Metallic behaviors are observed with the residual resistivity as low as ${0.12\ \mu\Omega\cdot\rm cm}$ at 0.4 K in zero field. The ratio of the room-temperature resistivity to the residual resistivity is about ${\rho_{\rm 300K}/\rho_{\rm 0.4K}\simeq100}$, suggesting a high quality of the measured single crystal. A sharp drop of the resistivity around $T_{\rm N}\simeq3$ K is observed, indicating that the scattering from the local moment fluctuations is greatly reduced in the AFM state.
In zero field below $T_N$, Fermi liquid-like behavior of $\rho=A\cdot{T^{\alpha}}+{\rho_{0}}$ and $\alpha\simeq2$ is observed. In the external field, the coefficient $A$ and the residual resistivity ${\rho_{0}}$ are strongly enhanced in the vicinity of the QCP. The Fermi liquid behavior with parabolic temperature dependence ($\alpha\simeq2$) gradually evolves into non-Fermi liquid-like behavior with linear temperature dependence ($\alpha\simeq1$) at the critical field (Fig.~\ref{fig:RT}(b-d)). These phenomena are similar to the heavy fermion system, when the material was tuned to the quantum critical region by the magnetic field~\cite{VMPNAS2019}. Although in~\TNA, Tm$^{3+}$ moments are rather localized and the Kondo temperature is much smaller compared to heavy fermion systems, the conduction electrons are still greatly scattered by the enhanced spin fluctuations in the vicinity of the QCP.

Further characterization with isothermal field-dependent resistivity is presented in Fig.~\ref{fig:RB}(a). Peak-like anomalies are observed at the transition field, and these peaks get sharper with decreasing temperature. At the lowest measured temperature 0.4 K, where the dominant contribution is from the residual resistivity, a clear maximum developed centered around 0.9 T, which again reflects the great enhancement of the spin fluctuation in the vicinity of the critical point. The contour plot of the resistivity is shown in Fig.~\ref{fig:RB}(b). Interestingly, this contour plot is identical to the field temperature-dependent magnetic entropy presented in Fig.~\ref{fig:Sm}(c). The Tm$^{3+}$ moments are well localized in \TNA, and the Kondo effect plays a small role here. This similarity between the resistivity and the magnetic entropy further confirms that the resistivity is mostly contributed by the spin fluctuation scattering of the Tm$^{3+}$ local moments.

\section{Conclusion}

In conclusion, single crystals of TmNi$_3$Al$_9$ with isolated two-dimensional honeycomb layers were synthesized. The magnetic properties were investigated using specific heat, magnetization, electrical transport, and neutron diffraction measurements. The experimental results, along with the CEF calculation, established an Ising-like moment of the Tm$^{3+}$ ion with the easy direction along the crystal $c$ axis. A long-range magnetic order was found below ${T_{\rm N}=3}$ K. The Tm$^{3+}$ moments are AFM coupled both within the $ab$~plane and along the $c$~axis. In external fields $B\parallel c$, the magnetic order was gradually suppressed around ${B_{c} \simeq 0.9}$~T. Strong spin fluctuations, induced as the system is tuned through the critical point, are evidenced by the broad peak in the magnetic susceptibility, the enhanced magnetic entropy, and the enhancement of the spin disorder scattering in the electrical resistivity measurements. However, the detailed spin dynamics remain unclear. Further investigations using inelastic neutron scattering techniques are needed, which may provide more information about the spin fluctuations and possible anisotropy of the exchange interactions in the 2D honeycomb lattice of the Tm$^{3+}$ Ising moments.

\begin{acknowledgments}

The research at SUSTech was supported by the National Natural Science Foundation of China (Grants No.~12134020, No.~11974157, No.~12174175 and No.~12104255), the Guangdong Basic and Applied Basic Research Foundation (Grant No.~2021B1515120015), and the National Key Research and Development Program of China (Grants No.~2021YFA1400400). This work was also supported by Shenzhen Key Laboratory of Advanced Quantum Functional Materials and Devices (Grant No.~ZDSYS20190902092905285), and the Shenzhen Science and Technology Program (Grant No.~KQTD20200820113047086).
The work is further supported by the Ministry of Science and Technology of China (Grants No.~2018YFE0103200 and No.~2021YFA1400300), by Shanghai Municipal Science and Technology Major Project (Grant No.~2019SHZDZX04), and by the Research Grants Council of Hong Kong with General Research Fund (Grant No.~17306520).
The Major Science and Technology Infrastructure Project of Material Genome Big-science Facilities Platform supported by Municipal Development and Reform Commission of Shenzhen.
Neutron diffraction measurements used resources at the Spallation Neutron Source, a DOE Office of Science User Facility operated by the Oak Ridge National Laboratory.

\end{acknowledgments}

\end{document}